\documentstyle[12pt]{article}
\begin{document}
\title{An Operator Formulation of Classical Mechanics and Semiclassical Limit}
\author{Slobodan Prvanovi\'c and Zvonko Mari\'c \\ Institute of Physics, P.O. Box 57, 11001 Belgrade, \\ Serbia}
\maketitle
\begin{abstract}
The generalized $h$-dependent operator algebra is defined ($0\le h
\le h_o$). For $h=h_o$ it becomes equivalent to the quantum
mechanical algebra of observables and for $h=0$ it is equivalent to
the classical one. We show this by proposing how the main features of
both mechanics can be defined in operator form.
\end{abstract}

The present considerations will address a part of kinematical aspect of  quantum mechanics (QM) and classical mechanics (CM) in order to investigate the correspondence principle. That is, the semiclassical limit of QM will be discussed in the algebraic framework. The dynamics will be just mentioned here. For this purpose an operator formulation of classical mechanics is proposed. It is very similar to those exposed and used in [1-4]. Our intention is to find such a formulation in which the main characteristics of QM and CM can be preserved. More precisely, for the new formulation the following should hold: 1.) the observables and states are in the 1-1 correspondence with the adequate ones of the standard formulation of QM and CM, 2.) the commutation relations among observables and the relations among the eigenstates of observables are the same as are those in the standard formulations and 3.) the mean values are not altered. The mathematical arena that will be used can be seen as a direct product of coordinate and
momentum representations of QM, so it will mimic phase space formulation of CM.

Firstly, we introduce the generalized operator $h$-dependent
algebra of observables which is defined as the algebra of
polynomials in operators $\tilde Q$, $\tilde P$ and $\hat{\cal I}$ (with real coefficients) that are defined as
\begin{equation}
 \tilde Q = \hat Q
\otimes \hat I \otimes \left[\hat R_q + \left(1-{h\over
h_o}\right)\hat R_p\right] + \hat I \otimes \hat Q \otimes \hat
R_p ,
\end{equation}
\begin{equation}
 \tilde P = \hat P \otimes \hat I \otimes \hat R_q +
\hat I \otimes \hat P \otimes \left[\left(1-{h\over h_o}\right)
\hat R_q + \hat R_p \right] ,
\end{equation}
and $\hat{\cal I}=\hat I \otimes \hat I \otimes \hat I$. These operators act in ${\cal H}_q \otimes {\cal H}_p \otimes {\cal H}_r$ where $q$, $p$ and $r$ are just indices. The first two spaces are the rigged Hilbert spaces and the third is a two-dimensional Hilbert space. More concretely, ${\cal H}_q$ and
${\cal H}_p$ are formally identical to the rigged Hilbert space of states
which is used in nonrelativistic QM of a single particle with the
one degree of freedom when spin is neglected. The indices $q$ and
$p$ serve only to denote that the choice of a basis in these
spaces is {\sl a priori} fixed when the semiclassical limit is under considerations. For the basis in ${\cal H}_q \otimes
{\cal H}_p$ we take $\vert q \rangle \otimes \vert p \rangle$.
Here $\vert q \rangle$ and $\vert p \rangle$ are the eigenvectors of
$\hat Q$ and $\hat P$, respectively. Then, ${\cal H}_q \otimes
{\cal H}_p $ can be seen as an analogue of the phase space. The
third space is introduced only for the formal reasons. The parameter
$h$ takes values from $0$ to $h_o$, where $h_o$ is related to QM
(the nonvanishing Planck constant) while for $h=0$ the
above algebra will be related to CM. The operators $\hat Q$ and
$\hat P$ are as the operators representing coordinate and momentum
in standard QM: they do not commute ($[\hat Q,\hat P]=i\hbar\hat I$),
they are Hermitian, {\sl etc.}

For the projectors
$\hat R_q$ and $\hat R_p$ the following relations should hold:
$\hat R_q \hat R_p =0$, $\hat R_q \hat R_q =\hat R_q$, $\hat R_p
\hat R_p = \hat R_p$, $\hat R_q ^\dagger =\hat R_q$, $\hat R_p ^
\dagger =\hat R_p$ and $\hat R_q +\hat R_p =\hat I$. They have no physical meaning and are introduced to ensure desired  properties of the polynomials in  $\tilde Q$ and $\tilde P$ for the extreme values of $h$.

When the above algebra of operators is represented with respect to the basis $\vert q \rangle \otimes \vert p \rangle \otimes \vert r_i \rangle$, where $i=\{ q,p\}$ and $\vert r_i \rangle$ is the eigenvector for $\hat R_i$ ($\langle r_i \vert r_j \rangle = \delta _{i,j}$), for $h=h_o$, it becomes equivalent to the representation (with respect to the same basis) of

\begin{equation}
\hat Q_{qm} = \hat Q \otimes \hat I \otimes \hat R_q + \hat I \otimes
\hat Q \otimes \hat R_p ,
\end{equation}
\begin{equation}
\hat P_{qm} = \hat P \otimes \hat I \otimes \hat R_q + \hat I \otimes
\hat P \otimes \hat R_p .
\end{equation}
This algebra and the appropriate eigenvectors are in the one-to-one correspondence with the standard formulation of QM (defined in a single rigged Hilbert space). For example, it holds: $[\hat Q_{qm},\hat  P_{qm}]=i\hbar\hat{\cal I}$, as it is necessary. On the other hand, due to the mentioned properties of $\hat R_q$ and $\hat R_p$, the standard representation of QM observable, {\sl e.g.}, $h(\hat Q,\hat P)$, is now translated to
$$
h(\hat Q_{qm},\hat P_{qm})=f(\hat Q,\hat P)\otimes\hat I \otimes\hat R_q +
\hat I \otimes f(\hat Q,\hat P)\otimes\hat R_p.
$$
If $\vert\Psi_i\rangle$ were eigenstates of $h(\hat Q,\hat P)$, then
$$
\vert \tilde \Psi_i\rangle=c_q\vert\Psi_i\rangle\otimes\vert a\rangle
\otimes\vert r_q\rangle+
c_p\vert b\rangle\otimes\vert\Psi_i\rangle\otimes\vert r_p\rangle,
$$
are eigenstates of $h(\hat Q_{qm},\hat P_{qm})$ with the same
eigenvalues if the coefficients $c_q$ and $c_p$ satisfy
the condition $\vert c_q\vert^2+\vert c_p\vert^2=1$ and if the vectors
$\vert a\rangle$ and $\vert b\rangle$, that are fixed at the beginning
of all considerations being arbitrarily picked, are normalized $\langle a \vert a \rangle = \langle b \vert b \rangle = 1$. The mean values and all the relations among eigenstates of the same or different observables are as in the standard formulation of QM (which can be easily seen).

On the other hand, for the above representation of $\tilde Q$ and $\tilde P$, but for $h=0$, the algebra becomes equivalent to the representation of
\begin{equation}
\hat Q_{cm} = \hat Q \otimes \hat I \otimes \hat I ,
\end{equation}
\begin{equation}
\hat P_{cm} = \hat I \otimes \hat P \otimes \hat I .
\end{equation}
This algebra and the appropriate eigenstates are in the 1-1 correspondence with the standard formulation of CM (defined in the phase space). Namely, to the $c$-number formulation of a CM observable, {\sl e.g.}, $h(q,p)$, now corresponds $h(\hat Q_{cm},\hat P_{cm}) = h( \hat Q \otimes \hat I , \hat I \otimes \hat P ) \otimes \hat I$. Such an algebra is manifestly a commutative
one. The vectors $\vert q_o\rangle\otimes\vert p_o\rangle\otimes
(c_q\vert r_q\rangle+c_p\vert r_p\rangle)$ are eigenstates of all
CM observables (with the eigenvalues $h(q_o,p_o)$). These vectors are
the analogs of the points in phase space for a CM system with
one degree of freedom. For these pure states it holds:
\begin{eqnarray}
&\vert q_o\rangle\langle q_o\vert\otimes\vert p_o\rangle\langle p_o\vert
\otimes
(c_q\vert r_q\rangle+c_p\vert r_p\rangle )
(c^*_q\langle r_q\vert+c^*_p\langle r_p\vert )=& \nonumber \\
=&\smallint\smallint\delta(q-q_o)\delta(p-p_o)
\vert q\rangle\langle q\vert\otimes\vert p\rangle\langle p\vert dqdp
\otimes
(c_q\vert r_q\rangle+c_p\vert r_p\rangle )
(c^*_q\langle r_q\vert+c^*_p\langle r_p\vert )&= \nonumber \\
&=\delta(\hat Q-q_o)\otimes \delta(\hat P-p_o)\otimes
(c_q\vert r_q\rangle+c_p\vert r_p\rangle )
(c^*_q\langle r_q\vert+c^*_p\langle r_p\vert ).&
\nonumber
\end{eqnarray}

Guided by this, the mixed CM states now can be defined as
$\rho(\hat Q\otimes\hat I,\hat I\otimes\hat P)\otimes
(c_q\vert r_q\rangle+c_p\vert r_p\rangle)
(c^*_q\langle r_q\vert+c^*_p\langle r_p\vert)$.
All CM states will be Hermitian, non-negative operators and normalized
to $\delta^2(0)$ if for $\rho(q,p)$ it holds that: $\rho (q,p) \in {\bf R}$,  $\rho (q,p) \geq 0$ and $\int \int \rho (q,p) dqdp = 1$ as in the standard phase space formulation of CM.
The mean values of both QM and CM observables are now calculated by the Ansatz: $\langle\hat A\rangle=Tr(\hat\rho\hat A)/Tr\hat\rho$,
so the norm $\delta^2(0)$ does not affect anything in the proposal. This, and the fact that the phase space formulation of CM appears through the kernels of the operator formulation in the $\vert q \rangle \otimes \vert p \rangle \otimes \vert r_i \rangle$ representation, can be used as the proof of equivalence of these two formulations.

There will be a complete correspondence between the $c$-number
formulation and the above given operator formulation of CM if the
dynamical equation is defined as the Liouville equation, where the
partial derivations within the Poisson bracket are done with
respect to the operators $\hat Q_{cm}$ and $\hat P_{cm}$. The
dynamical equation of QM representatives should be the
Schr\"odinger (von Neumann) equation as it is in the standard
formulation of QM.

The Hamilton function $h(q,p)$ of CM and the
Hamiltonian $h(\hat Q , \hat P )$  of QM (if they are addressing the
same physical system, for example the harmonic oscillator) are
represented here by $h(\hat Q _{cm} , \hat P _{cm} )$ and $h(\hat
Q _{qm} , \hat P _{qm})$, respectively. The last two operators
follow from $h(\tilde Q , \tilde P )$ for $h=0$ and $h=h_o$. (It
is understood that one should work in $\vert q \rangle \otimes
\vert p \rangle \otimes \vert r_i \rangle$ representation. This we
have not proceeded here  only for the sake of simplicity of
expressions.) Therefore, the semiclassical limit of QM (or the correspondence principle to be more accurate) can be established through the
generalized operator algebra, since for the one extreme value of
$h$ it expresses QM properties while for the other value of $h$ it
has CM ones. This holds for each polynomial with real coefficients
in coordinate and momentum no matter of how these operators are
ordered. (The ordering problem we shall discuss elsewhere.) The states are seen as secondary in the present proposal. That is, the meaningful states are solutions of the appropriate eigenvalue problems and, because QM and CM
observables are essentially different, they differ, too. At this
place it should be remarked that since the purpose of the
introduced framework was to discuss the semiclassical limit of QM
without altering the most important features of both mechanics, it
was necessary to take ${\cal H}_q \otimes {\cal H}_p \otimes{\cal
H}_r$ which is much wider than ${\cal H}$, where quantum mechanics
is irreducible represented. Only a subspace of ${\cal H}_q \otimes
{\cal H}_p \otimes{\cal H}_r$, that is formed over the basis
$\vert\tilde\Psi_i\rangle$, has the QM interpretation. It depends
on the choice of $\vert a\rangle$, $\vert b\rangle$, $c_q$ and
$c_p$ which, after being initially fixed, give the irreducible
representation of QM.

\begin{enumerate}

\item
T.N. Sherry and E.C.G. Sudarshan, Phys. Rev. D {\bf 18} 4580 (1978).
\item
T.N. Sherry and E.C.G. Sudarshan, Phys. Rev. D {\bf 20} 857 (1979).
\item
S.R. Gautam {\sl et al.}, Phys. Rev. D {\bf 20} 3081 (1979).
\item
S. Prvanovi\'c and Z. Mari\'c, Toward the collapse of state,  quant-ph/9910020, submitted for publication in Phys. Rev. A
\end{enumerate}

\end{document}